\documentclass[journal]{IEEEtran}
\usepackage{cite}
\usepackage{amsmath,amssymb,amsfonts}
\usepackage{algorithmic}
\usepackage{graphicx}
\usepackage{float}
\usepackage{algorithm,algorithmic}
\usepackage{hyperref}
\hypersetup{hidelinks=true}
\usepackage{textcomp}
\usepackage{booktabs,ragged2e}
\usepackage[flushleft]{threeparttable}

\usepackage{xr-hyper}
\usepackage{hyperref}
\usepackage{doi}
\usepackage{threeparttable}
\usepackage{tabularx}
\newcolumntype{L}{>{\raggedright\arraybackslash}X}
\def\BibTeX{{\rm B\kern-.05em{\sc i\kern-.025em b}\kern-.08em
    T\kern-.1667em\lower.7ex\hbox{E}\kern-.125emX}}
\begin{document}
\title{Quaternion-Based Predictive Framework for Scapulohumeral Coordination}
\author{Ondrej Zoufaly, Edward K. Chadwick, Dimitra Blana, Matej Daniel

\thanks{This work has been submitted to the IEEE for possible publication. Copyright may be transferred without notice, after which this version may no longer be accessible. This research was supported by the Czech Ministry of Education, Youth and Sports project No. CZ.02.01.01/00/23\_020/0008512 and Czech Grant Agency project No. 23-06920S.}
\thanks{O. Zoufaly is with Department of Mechanics, Biomechanics and Mechatronics, Faculty of Mechanical Engineering, Czech Technical University in Prague, Technicka 1902/4, Prague 6, 16000, Czech Republic (email: ondrej.zoufaly@fs.cvut.cz).}
\thanks{E. K. Chadwick is with University of Aberdeen, School of Engineering, 362 Fraser Noble Building, Old Aberdeen Campus, Elphinstone Road, Aberdeen, AB24 3UE, United Kingdom (email: edward.chadwick@abdn.ac.uk).}
\thanks{D. Blana is with University of Aberdeen, School of Medicine, Medical Sciences and Nutrition, Polwarth Building, Foresterhill Rd, Aberdeen, AB25 2ZD, United Kingdom (email: dimitra.blana1@abdn.ac.uk).}
\thanks{M. Daniel is with Department of Mechanics, Biomechanics and Mechatronics, Faculty of Mechanical Engineering, Czech Technical University in Prague, Technicka 1902/4, Prague 6, 16000, Czech Republic (email: matej.daniel@cvut.cz).}}

\maketitle

\begin{abstract}
Scapulohumeral rhythm (SHR), the coordinated motion between the scapula and humerus during arm elevation, is frequently altered in rotator cuff pathologies, yet the mechanical principles underlying coordination redistribution remain difficult to explain from experimental data alone. This study presents a predictive optimal control framework for investigating scapulohumeral coordination, combining a quaternion-based shoulder model with EMG-informed muscle parameter calibration. The quaternion formulation eliminated kinematic singularities and associated non-physiological activation artifacts observed in the Euler-angle model, while maintaining comparable tracking accuracy. EMG-informed calibration reduced discrepancies between predicted and measured muscle excitations by up to 60\% on independent validation tasks. In predictive simulations where only thoracohumeral elevation was prescribed, scapular and clavicular kinematics emerged from musculoskeletal mechanics and minimization of muscular effort, producing SHR values consistent with established experimental ranges. Simulated rotator cuff deficiency resulted in increased reliance on glenohumeral rotation. The proposed framework may serve as a tool for understanding impaired coordination patterns across a broad range of shoulder pathologies, with potential to inform personalized rehabilitation strategies and the design of assistive and prosthetic devices.

\end{abstract}

\begin{IEEEkeywords}
Shoulder, Quaternion, Optimal Control, Scapulohumeral rhythm, rotator cuff pathology
\end{IEEEkeywords}

\section{Introduction}
\label{sec,introduction}

Coordinated motion of the shoulder girdle joints is fundamental to functional upper-limb movement. During arm elevation, the scapula and humerus interact through a dynamically regulated scapulohumeral rhythm (SHR) \cite{inman_observations_1996,bagg_biomechanical_1988,mcclure_direct_2001,mcquade_dynamic_1998}, which distributes motion and loading across the shoulder complex. Alterations in this coordination are frequently observed in rotator cuff (RC) pathology with full-thickness tears affecting over 20\% of the general population \cite{yamamoto_prevalence_2010,minagawa_prevalence_2013}. In RC pathology, most experimental studies have reported decreased SHR with increased scapulothoracic contribution during arm elevation, suggesting long-term neuromuscular compensation for glenohumeral (GH) dysfunction \cite{ueda_comparison_2019,paletta_shoulder_1997}. In contrast, acute reduction of pain via subacromial injection in patients with full-thickness tears has been associated with increased SHR, reflecting greater GH contribution when pain-driven inhibition is removed \cite{scibek_shoulder_2008}. Furthermore, pain itself has been identified as an independent contributor to SHR alteration \cite{scibek_rotator_2009}. These findings suggest that the mechanical consequences of structural cuff deficiency and the neuromuscular adaptations driven by pain produce distinct coordination strategies. Understanding the mechanical principles governing scapulohumeral coordination is therefore essential for interpreting shoulder dysfunction.

Experimental quantification of shoulder kinematics and muscle coordination remains challenging. Scapular motion is particularly difficult to measure accurately using skin-mounted markers due to soft-tissue artifacts and anatomical constraints that limit direct tracking of the scapula \cite{karduna_dynamic_2000, mcclure_direct_2001}. Electromyography (EMG) provides insight into muscle activation patterns but is subject to cross-talk, normalization uncertainty, and limited ability to directly measure muscle force production \cite{luca_use_1997, buchanan_neuromusculoskeletal_2004}. Furthermore, direct measurement of GH joint contact forces is restricted to instrumented implants and therefore available only in limited populations \cite{bergmann_vivo_2011}. These constraints motivate the development of computational frameworks capable of estimating muscle forces, joint loading, and coordination strategies under controlled mechanical assumptions.

State-of-the-art shoulder musculoskeletal models have provided important insights into GH stability \cite{belli_does_2023, daniel_scapulohumeral_2026}, muscle coordination \cite{seth_muscle_2019}, and joint loading during arm elevation \cite{hasan_modeling_2025}. However, most existing implementations rely on forward-dynamics simulations or trajectory-tracking formulations that reproduce experimentally measured motion. Although these approaches yield valuable descriptive information regarding muscle forces and joint loads, they inherently constrain scapular kinematics to prescribed trajectories and therefore limit investigation of emergent coordination patterns and mechanical redistribution under altered muscle capacity. Recent studies have applied optimal control to the shoulder with reduced degrees of freedom \cite{hamad_opensim_2026, attias_musculoskeletal_2026}, but none have addressed predictive SHR with the full shoulder girdle model. Recently, an optimal control approach with the Delft shoulder model was used to predict scapulothoracic kinematic changes under simulated scapular stabilizer weakness \cite{russell_shoulder_2026}. However, their approach relied on Euler-angle kinematics which limited sagittal plane simulations due to instabilities. 

Predictive simulations of SHR derived from optimal control formulations would enable the investigation of coordination strategies that arise directly from musculoskeletal mechanics and physiological objectives. In lower-limb biomechanics, predictive optimal control approaches have successfully reproduced walking gait \cite{dorschky_estimation_2019, van_den_bogert_implicit_2011, de_groote_evaluation_2016, afschrift_benchmarking_2025, falisse_rapid_2019} and running \cite{nitschke_efficient_2020} patterns, and have been used to study coordination adaptations in pathological conditions \cite{van_den_bosch_estimating_2025,falisse_physics-based_2020,febrer-nafria_evaluation_2022} or after amputation \cite{vandenberg_feasibility_2026,wang_optimal_2026}. These studies demonstrate the ability of predictive frameworks to capture emergent movement strategies without prescribing joint trajectories.

Extending predictive optimal control to the shoulder presents distinct challenges. The shoulder girdle comprises multiple rotational degrees of freedom spanning the sternoclavicular (SC), acromioclavicular (AC), and GH joints. Euler-angle–based parameterizations, commonly used in musculoskeletal modeling, are inherently prone to kinematic singularity called gimbal lock, which occur when two terminal rotational axes align \cite{clark_tracking_2020, creveaux_rotation_2018, phadke_comparison_2011, senk_rotation_2006}. Near singular configurations, the mapping between generalized speeds and angular velocity becomes ill-conditioned \cite{hemingway_perspectives_2018}, potentially introducing numerical sensitivity and non-physiological behavior in forward or optimal control simulations. Alternative rotation representations have been proposed to overcome these limitations. Helical (screw) axis descriptions provide a compact geometric interpretation of three-dimensional rotation \cite{woltring_3-d_1994}, while unit quaternions offer a globally non-singular parameterization \cite{zoufaly_calculation_2025, shoemake_animating_1985} widely used in robotics, aerospace attitude control, and rigid-body dynamics \cite{guerrero-sanchez_quadrotor_2017,islam_model_2019,reyes-valeria_lqr_2013}.

While advances in kinematic representation improve numerical robustness, predictive simulations remain highly sensitive to underlying muscle–tendon parameters. Most musculoskeletal shoulder models rely on generic anatomical datasets derived from cadaveric measurements \cite{van_der_helm_geometry_1992, asadi_nikooyan_development_2011, holzbaur_model_2005}, which may not accurately reflect subject-specific strength or fiber optimal length. In inverse-dynamics or trajectory-tracking simulations, such discrepancies are partially mitigated by prescribed kinematics. In contrast, predictive optimal control formulations are more sensitive to parameter inaccuracies, as muscle properties directly influence emergent coordination patterns and joint loading. Small deviations in maximum isometric force or optimal fiber length can substantially alter predicted activation strategies, particularly for scapular stabilizers operating near passive regions of their force–length curves. Subject-specific calibration of muscle parameters has therefore been proposed as a means of improving physiological consistency between simulated activations and experimentally measured EMG \cite{sartori_emg-driven_2012,luis_experiment-guided_2024,modenese_estimation_2016,valente_are_2014}. Incorporating EMG-informed adjustments within a predictive optimal control framework may enhance the physiological consistency of predicted neuromuscular activation and scapulohumeral coordination patterns.
  
The framework aims to (1) improve numerical robustness in multi-planar shoulder simulations by adopting a quaternion-based joint representation, (2) enhance physiological consistency between predicted and experimentally measured muscle activation patterns through EMG-informed calibration of muscle parameters, and (3) predict emergent scapulohumeral coordination and GH loading from musculoskeletal mechanical demands and physiological cost minimization, without prescribing scapular or clavicular kinematics. To illustrate the framework's capacity to investigate coordination redistribution under altered muscle conditions, simulations are performed for both a healthy shoulder and a simulated RC-limited condition in a single representative participant. This implementation therefore serves as a methodological proof-of-concept demonstrating the feasibility of predictive SHR simulations within a full shoulder girdle model. For methodological comparison, an Euler-angle formulation of the shoulder model was also implemented to illustrate the numerical limitations associated with kinematic singularities; however, all predictive simulations presented in this study were performed using the quaternion-based formulation.

\section{Methods}
\label{sec,methods}

\subsection{Quaternion representation of joint rotations}

A unit quaternion $ \mathbf{Q} \in \mathbb{R}^4 $ was used to represent three-dimensional joint rotations. The quaternion is defined as \cite{shoemake_animating_1985}

\begin{equation}
\mathbf{Q} =
\begin{bmatrix}
Q_0 \\
\mathbf{Q}_v
\end{bmatrix}
=
\begin{bmatrix}
Q_0 \\
Q_1 \\
Q_2 \\
Q_3
\end{bmatrix},
\end{equation}

where $Q_0$ denotes the scalar component and $\mathbf{Q}_v = [Q_1 \; Q_2 \; Q_3]^T$ denotes the vector component. Unit quaternions satisfy the normalization constraint

\begin{equation}
\|\mathbf{Q}\|^2 = Q_0^2 + Q_1^2 + Q_2^2 + Q_3^2 = 1.
\end{equation}

The relation between quaternion time derivative and angular velocity $\boldsymbol{\omega} \in \mathbb{R}^3$ is \cite{graf_quaternions_2008}:

\begin{equation}
\label{eq, quaternion kinematic mapping}
\dot{\mathbf{Q}} = \frac{1}{2}
\begin{bmatrix}
- \mathbf{Q}_v^T \\
Q_0 \mathbf{I}_3 + [\mathbf{Q}_v]_\times
\end{bmatrix}
\boldsymbol{\omega},
\end{equation}

where $[\mathbf{Q}_v]_\times$ denotes the skew-symmetric matrix associated with $\mathbf{Q}_v$. 

This formulation provides a globally nonsingular parameterization of three-dimensional rotations and was used exclusively for the dynamic simulation.

\subsection{Experimental data}
\label{subsec,experimental data}

Three healthy adults (2 males, 1 female; age: 37.0 $\pm$ 12.7 years; height: 177.3 $\pm$ 6.2 cm; mass: 83.0 $\pm$ 12.8 kg) with no history of shoulder pathology or upper-limb injury participated in the study. All participants provided written informed consent. Ethical approval was granted by the University of Aberdeen Physical Sciences and Engineering Ethics Board under Application No. 1290201.

Each participant performed three standardized arm elevation tasks: 90$^\circ$ of thoracohumeral elevation in the frontal, scapular, and sagittal planes. Each elevation plane was recorded in a separate trial. In addition, four activities of daily living (ADLs) were recorded: a simulated driving steering motion, drinking from a cup, reaching to a shelf, and lifting a 5kg vertically from the neutral position. All tasks were performed at self-selected speed. The elevation tasks were used for trajectory tracking (Section~\ref{subsec,IK tracking}), EMG-informed calibration (Section~\ref{subsec,calibration}), and SHR prediction (Section~\ref{subsec,SHR prediction}). The ADL tasks were used exclusively for validation of the calibrated muscle parameters.

Upper limb kinematics were recorded using an OptiTrack motion capture system (NaturalPoint, USA) at a sampling rate of 100 Hz and synchronized with EMG acquisition via MotionMonitor software (Innovative Sports Training, USA). Marker clusters were placed on the thorax, acromion (scapular tracking cluster), humerus, and forearm. Anatomical landmarks were digitized in a static standing calibration trial using a stylus. The identified landmarks consistent with the virtual model were associated with the corresponding marker clusters and used for subject-specific model scaling and inverse kinematics.

Surface EMG signals were recorded using a Delsys system (Delsys Inc., USA) at a sampling rate of 2000 Hz. Electrodes were placed over the upper and middle trapezius, anterior, lateral, and posterior deltoid, infraspinatus, and serratus anterior following standard placement guidelines \cite{hermens_development_2000}. Prior to experimental trials, maximum voluntary contractions (MVCs) were collected for each muscle and used to normalize EMG amplitude to maximum voluntary activation \cite{boettcher_standard_2008}.

Raw EMG signals were band-pass filtered (10–400 Hz) using a 2nd-order zero-lag Butterworth filter, full-wave rectified, and subsequently low-pass filtered at 4 Hz using a 2nd-order zero-lag Butterworth filter to obtain the linear envelope \cite{winter_biomechanics_2009}. The processed EMG envelopes were resampled to 100 Hz to match the kinematic sampling frequency and aligned to the inverse kinematics time base prior to simulation.

\subsection{Shoulder musculoskeletal model}

The shoulder musculoskeletal model developed by Chadwick et al. \cite{chadwick_real-time_2014} was adopted and modified for this study. The original model represents the SC, AC, and GH joints using Euler-angle parameterizations with YZX, YZX, and YZY rotation sequences, respectively. These correspond to clavicular protraction/retraction, elevation, and axial rotation at the SC joint; scapular internal/external rotation, upward/downward rotation, and anterior/posterior tilting at the AC joint; and humeral plane of elevation, elevation, and axial rotation at the GH joint. Elbow flexion/extension and forearm pronation/supination are represented by revolute joints. In its original form, the model comprises 11 degrees of freedom (DoFs) and 136 muscle elements. To reduce model complexity and focus on shoulder girdle kinematics, forearm pronation and supination were locked, resulting in a 10-DoF model with 126 active muscle elements.

Model scaling and inverse kinematics were performed in OpenSim \cite{delp_opensim_2007} using experimentally measured marker trajectories. Muscle–tendon paths and optimal fiber lengths were initially scaled using the standard OpenSim scaling procedure. For selected muscle elements (rhomboideus and triceps long head), optimal fiber lengths were subsequently adjusted within physiologically plausible ranges to avoid excessive passive force generation.

To enable numerically robust simulation of large three-dimensional shoulder rotations, the Euler-angle parameterizations of the SC, AC, and GH joints were reformulated using unit quaternions during the equations-of-motion formulation. The anatomical joint axes and rotation definitions remained unchanged. The quaternion representation was used exclusively for dynamic simulation to eliminate singularities associated with Euler-angle kinematics and to improve numerical stability in predictive optimal control. For analysis and interpretation, the simulated joint kinematics were converted back to the original Euler-angle definitions.

\subsection{Muscle action modeling}

Muscle forces were modeled using a Hill-type formulation with rigid tendon assumption. Each muscle element generates force as a function of activation, normalized fiber length, and contraction velocity according to standard force–length and force–velocity relationships. Passive fiber force was included through a nonlinear elastic element \cite{thelen_adjustment_2003,de_groote_evaluation_2016}. Muscle activation dynamics was modeled using a first-order differential equation \cite{chadwick_real-time_2014}. A detailed description of the muscle model is provided in the Supplementary Material (SM).

Muscle–tendon lengths and moment arms were approximated using polynomial functions of the joint coordinates. Polynomial coefficients were obtained by fitting to muscle–tendon geometry exported from the OpenSim model over the physiological range of motion \cite{van_den_bogert_implicit_2011}. Moment arms mapping from Euler-angles and quaternions follows our previous formulation \cite{zoufaly_calculation_2025}, ensuring dynamical equivalence between the two kinematic representations. The stopping criterion of the algorithm for adding more polynomial terms was that the change of RMSE between the approximation and OpenSim values was less than 3\%.

Muscle forces were converted to joint-level contributions through approximated moment arms. In the Euler-angle formulation, muscle forces generate generalized torques directly associated with the joint coordinates. In the quaternion-based formulation, muscle forces generate spatial moments applied to the rotational equations of motion.

\begin{figure*}[t]
\centering
\includegraphics[width=\textwidth]{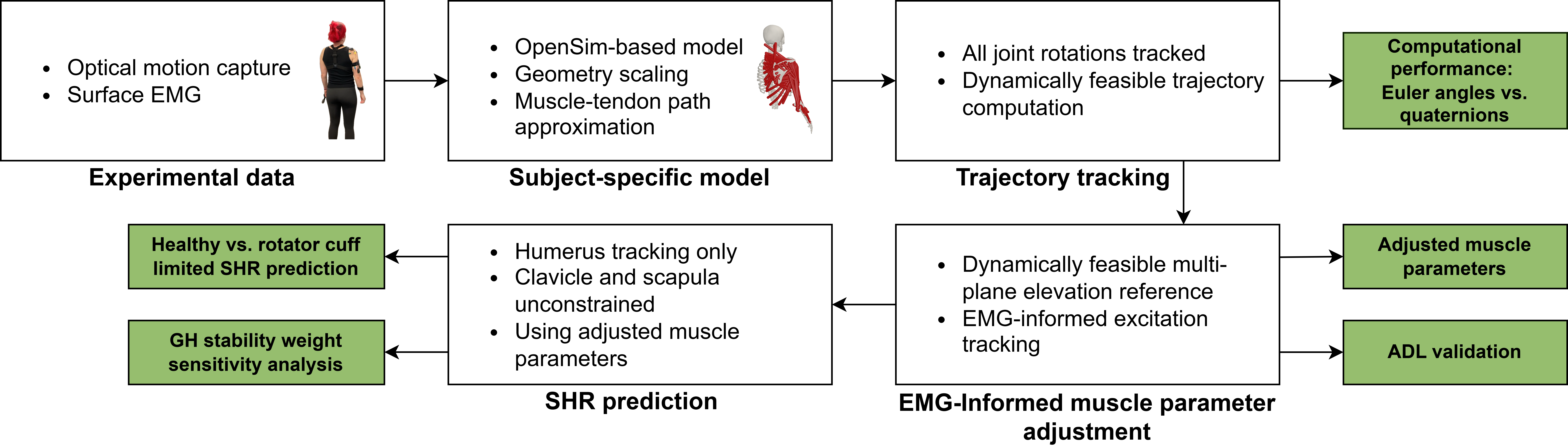}
\caption{Overview of the proposed modeling and optimization framework. Green boxes highlight outputs analyzed in the Results section.}
\label{fig,diag}
\end{figure*}

\subsection{Equations of motion}

Two sets of equations of motion were derived using SymPy Mechanics \cite{meurer_sympy_2017}. The first formulation employs Euler-angle parameterizations for spherical joints, while the second formulation uses unit quaternions. In both cases, the equations were generated symbolically and expressed in implicit form:

\begin{equation}\label{eq,eoms}
\mathbf{M}(\mathbf{q})\dot{\mathbf{u}} + \mathbf{f}(\mathbf{q}, \mathbf{u}) - \boldsymbol{\tau}_m = \mathbf{0},
\end{equation}

where $\mathbf{q} \in \mathbb{R}^{N_q}$ denotes the vector of generalized coordinates and $\mathbf{u} \in \mathbb{R}^{N_u}$ denotes the vector of generalized speeds. The matrix $\textbf{M}(\mathbf{q})$ is the generalized mass matrix, $\mathbf{f}(\mathbf{q},\mathbf{u})$ collects Coriolis and centrifugal effects, gravitational forces, passive joint and conoideum ligament contributions, and $\mathbf{\tau}_{m}$ represents muscle-induced generalized or spatial moments. Equation \eqref{eq,eoms} consists of $\mathbb{R}^{N_u}$ differential equations..

An additional set of $\mathbb{R}^{N_q}$ kinematic differential equations relates the time derivatives of the generalized coordinates $\dot{\mathbf{q}}$ to the generalized speeds $\mathbf{u}$. In the Euler-angle formulation, generalized speeds correspond directly to joint angles rates. In the quaternion formulation, generalized speeds correspond to angular velocities and are related to quaternion time derivative through the kinematic mapping described in (\ref{eq, quaternion kinematic mapping}).

In the Euler-angle formulation, the system comprises $N_q = 10$ and $N_u = 10$ (20 differential states). In the quaternion formulation, unit quaternions replace Euler angles for three spherical joints, yielding $N_q = 13$ and $N_u = 10$ (23 differential states).

The GH joint reaction force components in the corresponding optimization problem were included as an additional three variables, computed from three supplementary equilibrium equations at each collocation node in the implicit system.

\subsection{Optimal control formulation}

The shoulder simulation is formulated as a dynamical system in which the muscle activation dynamics and the multibody equations of motion are merged into a single set of first-order differential equations with state vector $\mathbf{x}$ defined as

\begin{equation}\label{eq,state vector}
\mathbf{x} = (\mathbf{q},\mathbf{u},\mathbf{a}) \in \mathbb{R}^{N_q + N_u + N_{mus}}
\end{equation}

where $\mathbf{a}$ denotes muscle activations, and input vector $\mathbf{e} \in \mathbb{R}^{N_{mus}}$, where $\mathbf{e}$ denotes muscle excitations. $\mathbf{q}$ and $\mathbf{u}$, as well as $N_q$ and $N_u$, depend on formulation.

To solve the optimal control problem, the continuous-time dynamical system described above was discretized through a direct collocation approach using Opty Python package\cite{moore_opty_2018}. In this method, the state and control trajectories are represented by discrete values and the system dynamics are enforced at each node as equality \cite{betts_practical_2010}. The midpoint integration scheme was used to approximate the dynamics between nodes, a fixed discretization of 0.04s was used for all simulations. Depending on the simulation objective, the cost function consists of a subset of physiologically and numerically motivated terms, which are detailed below.

\paragraph{Trajectory tracking}
\begin{equation}\label{eq,objective-traj}
\begin{split}
&J_{traj} = w_{traj} \int^{tf}_{t_{0}} \left( \sum^{N_{traj}}_{i=1} \left(error_{i}^{traj} \right)^2 \right)dt \\
\end{split}
\end{equation}

minimizes the error between experimentally measured segment orientations obtained from inverse kinematics and the corresponding simulated orientations. The orientations of scapula and humerus were measured in the world reference frame. The axial rotation of the clavicle (represented by rotation around $x$-axis) is difficult to capture, so it was allowed to rotate freely (its rotation is highly influenced by the presence of conoideum ligament as modeled in \cite{chadwick_real-time_2014}). In the Euler angles-based model free axial rotation of the clavicle was enabled by excluding the axial rotation from tracking. Free axial rotation of the clavicle in the quaternion-based model was modeled by tracking the acromioclavicular joint, which lies on the x-axis of the clavicle frame:

\begin{equation}\label{eq,x-axis point}
p^{world}_{AC} = \textbf{R}(\textbf{Q}^{SC}) p^{loc}_{AC}
\end{equation}

where $p_{AC}^C = (P_{AC},0,0)$ is the position of the acromioclavicular joint in the clavicle frame, $p_{AC}^W$ is the position of the acromioclavicular joint in the world reference frame and $\textbf{R}(\textbf{Q}^{SC})$ is the quaternion rotation matrix that maps the point from the clavicle frame to the world frame, where $\textbf{Q}^{SC}$ is the quaternion representing rotation of the clavicle in the world frame. This formulation constrains the orientation of the clavicle in space while leaving axial rotation unconstrained.

Scapula and humerus orientation error was computed as the difference between experimental and simulated joint angles for the Euler-angle–based model. For the quaternion-based model, the orientation error was defined by minimizing the vector part of the relative quaternion rotation error $\textbf{Q}\otimes \textbf{Q}^{-1}_{exp}$, which corresponds to minimizing the orientation mismatch in the tangent space. The rotation error of the elbow was measured in the humerus frame.

\paragraph{Muscle activation minimization}
\begin{equation}\label{eq,objective-act}
J_{act} = w_{act} \int^{tf}_{t_{0}} \left( \sum^{N_{mus}}_{j=1} a^2_{j} \right)dt
\end{equation}

minimizes the muscle activations squared.

\paragraph{Glenohumeral joint stability}
\begin{equation}\label{eq,objective-stab}
J_{\rm GH} = w_{GH} \int_{t_0}^{t_f}
\left[
0.31\left(\frac{R_{\rm sup\text{-}inf}}{R_{\rm comp}}\right)^2 +
\left(\frac{R_{\rm ant\text{-}pos}}{R_{\rm comp}}\right)^2
\right] dt
\end{equation}

minimizes destabilizing shear forces normalized by the compression force at the GH joint. Here, $R_{sup-inf}$, $R_{ant-pos}$ and $R_{comp}$ denote the superior–inferior, anterior–posterior, and compressive components of the GH joint reaction force, respectively. The reaction force components were expressed in the glenoid coordinate frame. The superior–inferior shear component was scaled by a factor of 0.31, obtained by normalizing the directional stability ratios reported in concavity–compression experiments \cite{lippitt_glenohumeral_1993}. The glenoid has 6.5$^\circ$ of inferior inclination and 13$^\circ$ of anterior version in the ISB reference frame (values are based on a segmented scapula STL model from The Visible Human Project \cite{ackerman_visible_1999}).

\paragraph{Experimental EMG tracking}

Processed EMG envelopes were treated as proxies for neural excitation and therefore compared to muscle excitation inputs rather than activation states.

\begin{equation}\label{eq,objective-emg}
J_{EMG} = w_{EMG} \int^{tf}_{t_{0}} \left( \sum^{N_{EMG}}_{k=1} \left(e_k - EMG_k \right)^2 \right)dt
\end{equation}

minimizes the difference between experimentally measured EMG and computed excitations of corresponding muscle elements.

\paragraph{Parameter regularization}

\begin{equation}\label{eq,objective-scaler}
J_{pars} = w_{pars} \sum^{N_{mus}^{scaled}}_{m=1} \Bigg( (scaler^m_{F_{max}}-1)^2 + (scaler^m_{l_{ce}^{opt}}-1)^2 \Bigg)
\end{equation}

is the term that keeps the scalers close to its generic value to restrict excessive scaling. $scaler^m_{F_{max}}$ scales the maximum isometric force of the $m$-th muscle group and $N_{mus}^{scaled}$ is the number of muscle groups being scaled. The same logic is applied for the optimal length $scaler^m_{l_{ce}^{opt}}$.

\paragraph{Numerical regularization and constraint stabilization}

To promote smooth solutions and improve numerical convergence, a quadratic regularization term penalizing the time derivative of the generalized speeds, excitations and reaction forces,respectively, was included \cite{falisse_rapid_2019}:

\begin{equation}
J_{\rm reg} = w_{reg}
\int_{t_0}^{t_f}
\Bigg(
\sum_{n=1}^{N_u} \dot{u}_n^2
+ \sum_{n=1}^{N_{mus}}  \dot{R}_n^2 \bigg)
+ w_{exc} \int_{t_0}^{t_f} \sum_{n=1}^{N_R} \dot{e}_n^2dt
\end{equation}

where $w_{exc}=10^{-3}$. In the Euler-angle formulation, the generalized speeds correspond to joint angle rates, such that penalizing $\dot{\mathbf{u}}^2$ minimizes generalized accelerations. In the quaternion formulation, generalized speeds correspond to angular velocities, and penalizing $\dot{\mathbf{u}}^2$ therefore minimizes angular accelerations. 

For the quaternion-based formulation, a hybrid constraint strategy was employed to ensure the unit quaternion across the simulation. The unit-norm condition for each quaternion was enforced as a hard algebraic equality constraint at the initial node ($t_0$):
\begin{equation}
\|\mathbf{Q}_j(t_0)\|^2 = 1, \quad j = 1, \dots, 3.
\end{equation}
For all subsequent collocation nodes, the unit-norm condition was enforced via the soft-penalty term in the objective function:

\begin{equation}
J_{\rm quat} = w_{quat}
\int_{t_0}^{t_f}
\Bigg(
\sum_{j=1}^{N_{quat}}
(\mathbf{Q}_j^T\mathbf{Q}_j - 1)^2
\Bigg) dt
\end{equation}

where the penalty weight $w_{quat}$ was set equal to the trajectory tracking weight $w_{traj}$. 

Depending on the simulation scenario, selected terms were activated or deactivated as summarized in Table \ref{tab,objective_terms} with corresponding weight values.

\begin{table}[t]
\centering
\begin{threeparttable}
\caption{Objective function terms and corresponding weights for each optimization problem. A dash indicates that the term was not active.}
\label{tab,objective_terms}
\small 
\setlength{\tabcolsep}{2pt} 
\begin{tabularx}{\columnwidth}{Lccccccc} 
\hline
\textbf{Optimization problem}  
& $J_{traj}$ & $J_{act}$ & $J_{GH}$ & $J_{EMG}$ & $J_{pars}$ & $J_{reg}$ & $J_{quat}$\\
\hline
Trajectory tracking  
& 200\tnote{a} & 1 & --- & --- & --- & $10^{-3}$ & 200 \\
\noalign{\smallskip} 
Muscle parameter optimization  
& 200 & 1 & 2 & 2 & 4 & $10^{-2}$ & 200 \\
\noalign{\smallskip}
SHR prediction (healthy)  
& 200\tnote{b} & 1 & 2 & --- & --- & $10^{-2}$ & 200 \\
\noalign{\smallskip}
SHR prediction (RC-limited)  
& 200\tnote{b} & 1 & 2--10\tnote{c} & --- & --- & $10^{-2}$ & 200 \\
\hline
\end{tabularx}
\begin{tablenotes} \footnotesize \item[a]The trajectory tracking weight corresponds to the quaternion-based formulation. In the Euler-angle comparison simulations, a weight of 50 was used to obtain comparable tracking strength due to differences in orientation error representation (see SM). \item[b]Only thoracohumeral rotation was tracked during SHR prediction; scapular and clavicular motions were predicted. \item[c]Sensitivity analysis was performed with $w_{GH}$ ranging from 2 (healthy value) to 10; see Section~\ref{subsec,GH sensitivity}. The value of 10 was used for the primary RC-limited results. \end{tablenotes}
\end{threeparttable}
\end{table}

\subsection{Trajectory tracking simulations}
\label{subsec,IK tracking}

Trajectory tracking simulations were performed to evaluate the ability of the Euler-angle–based and quaternion-based shoulder models to reproduce experimentally measured upper-limb motions. Three basic tasks were considered: arm elevation in the frontal, scapular, and sagittal planes. Each task was performed by three participants, resulting in nine tracking simulations for each model formulation.

In these simulations, the experimental joint kinematics obtained from inverse kinematics were tracked using the trajectory tracking cost term, together with muscle activation minimization and numerical regularization. The purpose of these simulations was not to quantify statistical differences between formulations, but to verify that the quaternion-based model achieves equivalent tracking accuracy while eliminating the kinematic singularities. Since tracking accuracy in both formulations depends on the relative weighting of the tracking term, a direct quantitative comparison of tracking performance would reflect weight selection rather than inherent formulation properties.

The Euler-angle formulation was therefore used only for methodological comparison to illustrate the numerical effects of kinematic singularities, while all subsequent predictive simulations presented in this study were performed using the quaternion-based representation.

\subsection{EMG-Informed muscle parameter adjustment}
\label{subsec,calibration}

To ensure consistency between simulated muscle activations and experimentally measured EMG signals, selected muscle parameters were adjusted using an EMG-informed calibration procedure. Specifically, multiplicative scalers were applied to the maximum isometric force ($F_{max}$) and optimal fiber length ($l_{ce}^{opt}$) of selected muscle groups. The muscle groups subject to scaling, as well as the mapping between experimental EMG channels and model muscle elements, are detailed in the SM.

To reduce the risk of overfitting to a single movement, the adjustment was performed using the three elevation trials concatenated into a single composite motion with 1 second pauses inserted between planes to ensure physiologically plausible resting configurations at transitions. The dynamically feasible trajectories obtained from the trajectory tracking simulations (Section~\ref{subsec,IK tracking}) were used as reference kinematics for the composite motion. By using dynamically consistent trajectories rather than raw inverse kinematics output, the calibration procedure was freed from residual kinematic errors arising from soft-tissue artifacts and scapulothoracic contact approximations, allowing the optimizer to focus on adjusting muscle parameters to minimize discrepancies between predicted excitations and measured EMG.

During calibration, the EMG tracking term was used to align simulated muscle excitations with measured EMG signals, together with the trajectory tracking, activation minimization, and numerical regularization terms, while the muscle parameter scalers were being adjusted. A quadratic regularization term constrained the scalers to remain close to their nominal values, limiting parameter deviation unless required to improve EMG consistency.

The adjusted muscle parameters were subsequently evaluated in validation simulations involving ADLs that were not included in the calibration process. The individualized parameters obtained from the calibration were then used in the SHR prediction simulations.

\subsection{Scapulohumeral rhythm prediction}
\label{subsec,SHR prediction}

Simulations of SHR prediction were performed to assess the model’s ability to predict scapular and clavicular kinematics when only thoracohumeral motion was prescribed. In these simulations, the thoracohumeral rotation obtained from the experimental inverse kinematics was tracked, while the motions of the scapula and clavicle were unconstrained. Scapular and clavicular kinematics therefore emerged from the dynamic interaction between muscle forces, passive structures, and the objective function.

The same composite elevation task described in Section~\ref{subsec,calibration} was selected for the prediction simulations. Arm elevation is widely used in experimental studies of SHR and provides a standardized and reproducible motion that enables comparison with existing literature.

The prediction simulations used muscle activation minimization and GH joint stability terms, together with numerical regularization, but excluded direct tracking of scapular and clavicular kinematics.

To demonstrate the potential of the optimal control framework for predicting kinematic adaptations under pathological conditions, both supraspinatus and infraspinatus were excluded to simulate a massive posterosuperior cuff deficiency. This severe condition was chosen deliberately to produce coordination changes of sufficient magnitude to clearly distinguish the predicted redistribution from numerical sensitivity or weighting artifacts. Excluding supraspinatus alone — while more representative of RC pathology — would produce smaller kinematic changes that could be difficult to differentiate from the inherent variability of the optimization. The goal was not to replicate a specific clinical tear pattern but to demonstrate the framework's capacity to predict substantial coordination redistribution in response to a major change in muscle capacity.

SHR was quantified as the ratio of incremental GH elevation to incremental scapulothoracic upward rotation, computed within three phases of thoracohumeral elevation: start–30$^\circ$, 30–60$^\circ$, 60–90$^\circ$, and start-90$^\circ$. Specifically, SHR was defined as $\Delta (TH-TS) / \Delta TS$, where $TH$ denotes thoracohumeral elevation and $TS$ denotes thoracoscapular upward rotation. A higher SHR value indicates greater GH contribution relative to scapular motion.

\begin{figure}[t]
\centering
\includegraphics[width=\columnwidth]{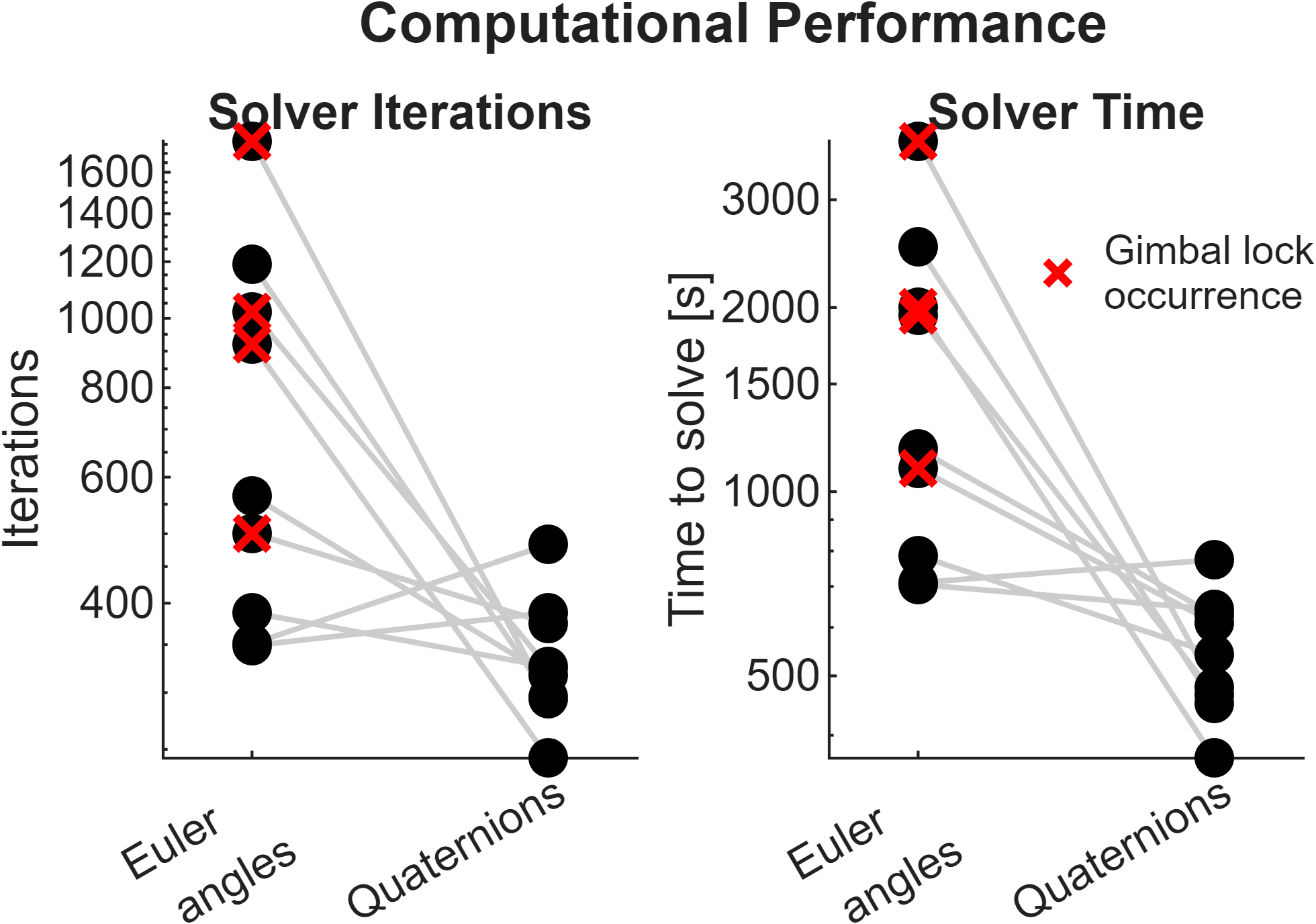}
\caption{Comparison of computational performance between Euler-angles and quaternion-based models during trajectory tracking simulations.}
\label{fig,comp_perf}
\end{figure}

\begin{figure}[t]
\centering
\includegraphics[width=\columnwidth]{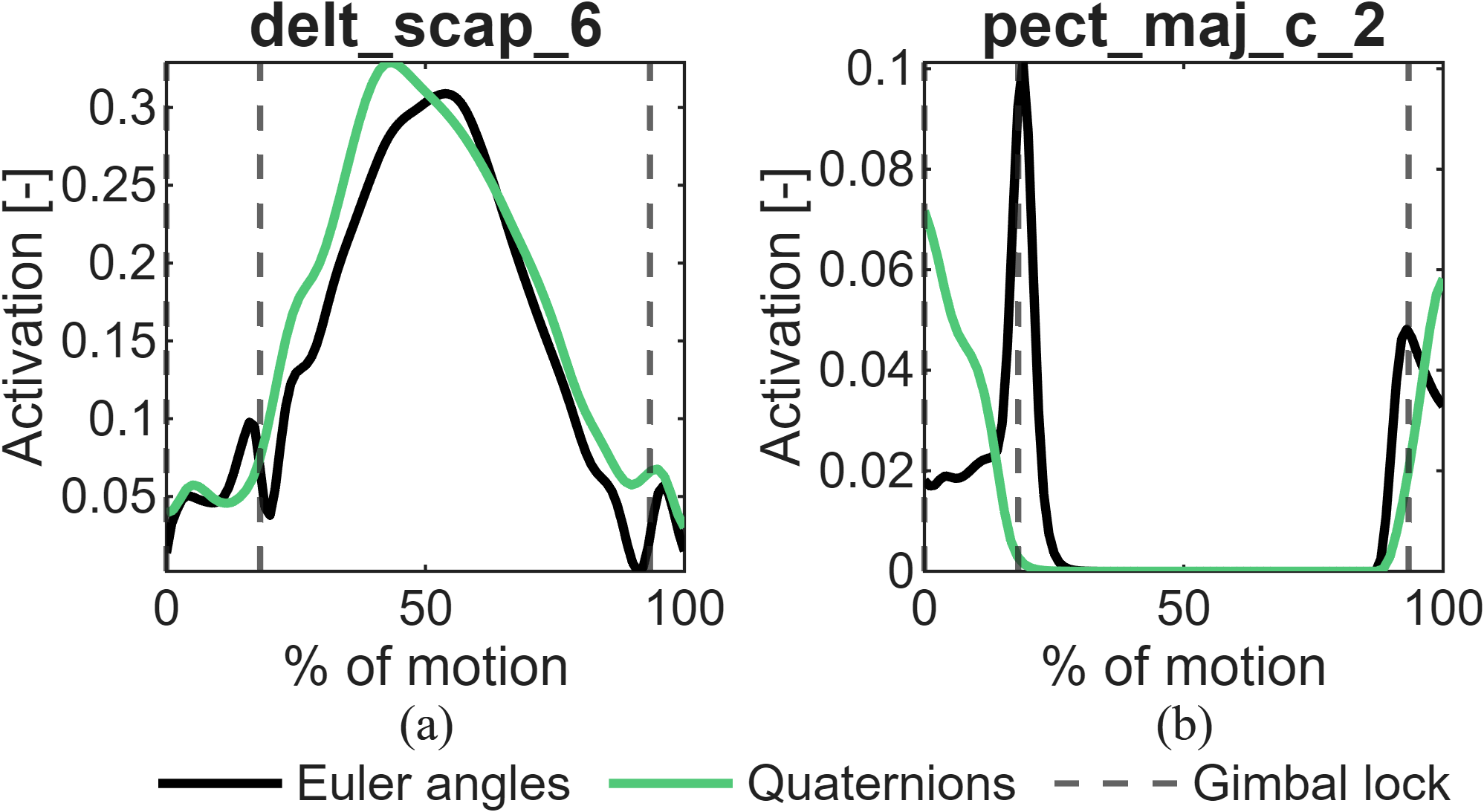}
\caption{Effect of gimbal lock on muscle activation in selected muscle elements in the Euler-angle–based model (black line) during humeral elevation in the frontal plane.}
\label{fig,activation_near_GL}
\end{figure}

\begin{figure}[t]
\centering
\includegraphics[width=\columnwidth]{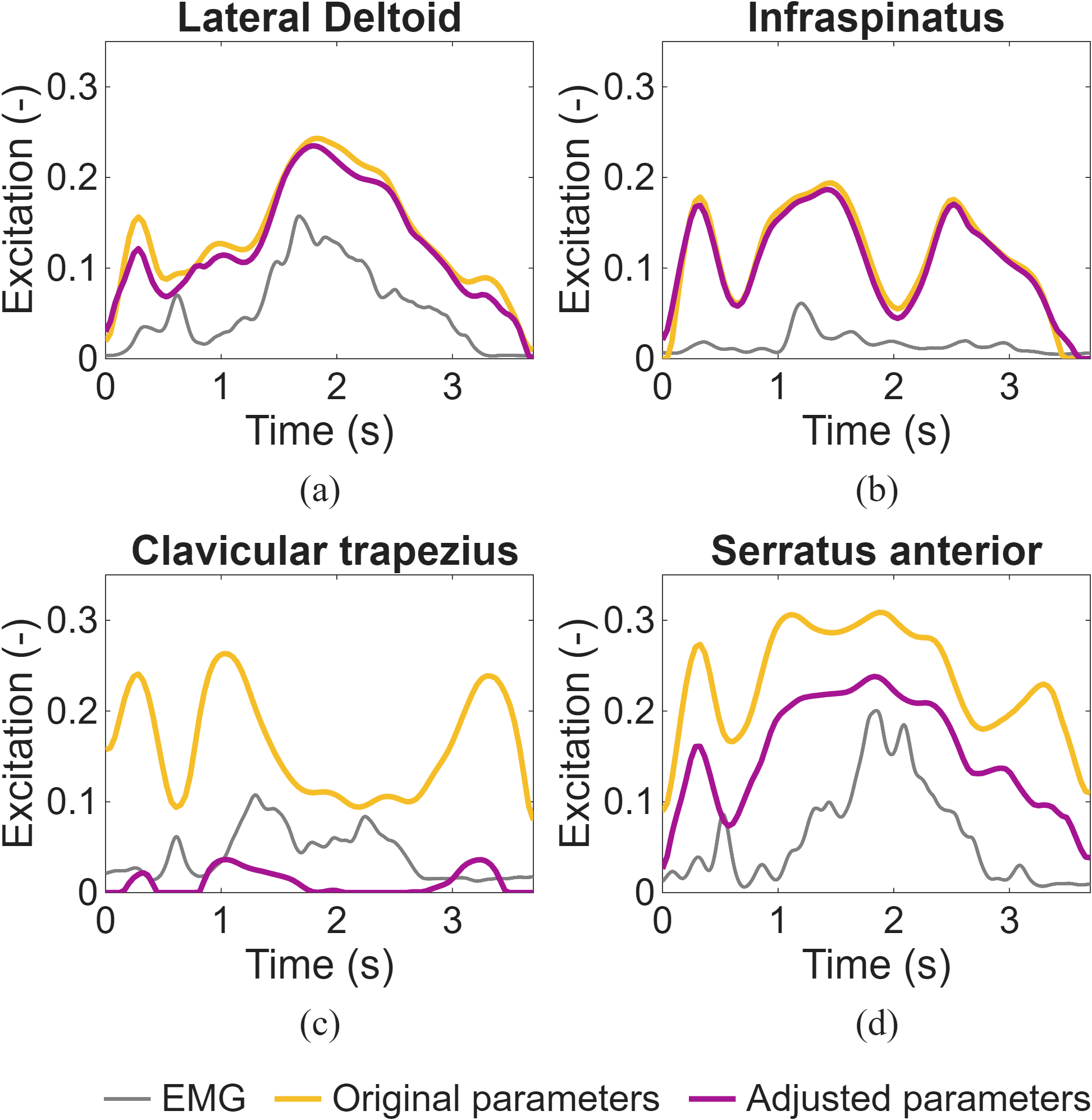}
\caption{EMG compared to the resulting muscle excitation with original and adjusted muscle paramaters during a shelf-reaching ADL.}
\label{fig,activation_validation}
\end{figure}

\begin{figure*}[t]
\centering
\centerline{\includegraphics[width=\textwidth]{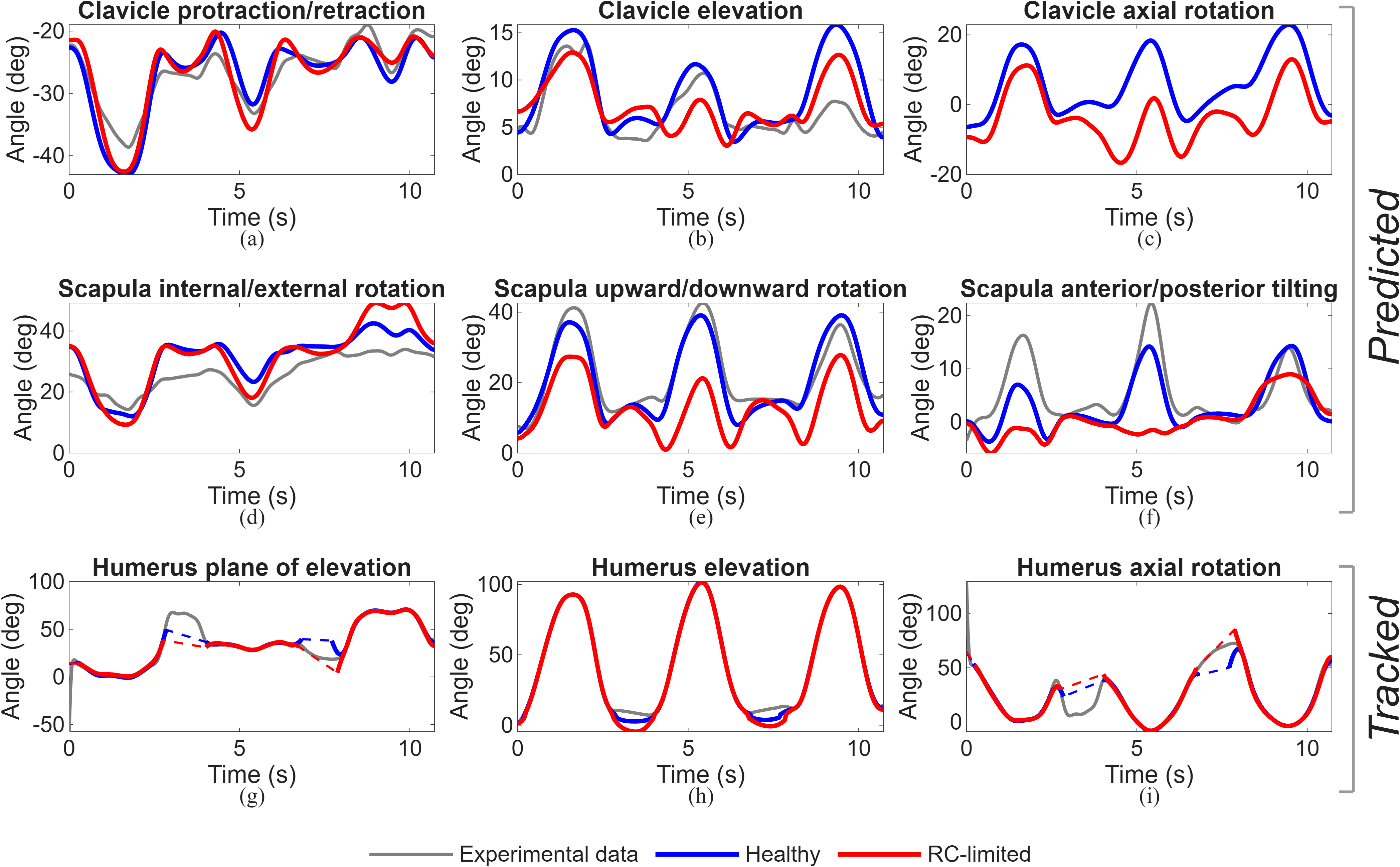}}
\caption{The resulting joint rotations from the SHR prediction of healthy condition (blue line) and RC-limited condition (red line) compared to the dynamically feasible reference trajectory calculated from experimental data in Section \ref{subsec,IK tracking} (grey line). Only humerus was tracked during these simulations. All angles are measured with respect to thorax. (g),(i) Due to gimbal lock in the YZY sequence of thoracohumeral rotation, values for the first and third axes were zeroed due to gimbal lock instability when elevation was below 8$^\circ$, corresponding segments in the kinematics plots are linearly interpolated and represented by dashed lines.}
\label{fig,kinematics_prediction}
\end{figure*}

\section{Results}

\subsection{Quaternions vs. Euler angles}
\label{sec:quat_vs_eul}

The root-mean-square error (RMSE) between simulated trajectories and experimental inverse kinematics across three participants and three elevation planes was comparable between formulations. For the clavicular (excluding axial rotation), RMSE was 1.90$^{\circ}$ for the Euler-angle model and 2.79$^{\circ}$ for the quaternion-based model, respectively. For thoracoscapular angles, RMSE was 2.72$^{\circ}$ and 3.67$^{\circ}$, and for thoracohumeral angles 0.61$^{\circ}$ and 4.26$^{\circ}$, for Euler-angles and quaternion formulations, respectively. Higher RMSE value for thoracohumeral angles in quaternion-based model is caused by the first and third axis instability when calculating corresponding Euler angles from quaternion. These results indicate similar tracking accuracy under full trajectory tracking conditions.

Computational performance for both formulations is summarized in Fig.~\ref{fig,comp_perf}. All simulations were solved as trajectory-tracking optimal control problems using identical objective terms and solver settings. Simulations in which the Euler-angle formulation encountered a gimbal lock configuration are highlighted.

Although the Euler-angle formulation produced feasible solutions in simulations that passed through kinematic singularities, the resulting muscle activation profiles exhibited abrupt, non-physiological spikes for muscles crossing the GH joint (Fig.~\ref{fig,activation_near_GL}). These artifacts were not associated with changes in tracking accuracy but originated from the singular kinematic representation. In contrast, the quaternion-based formulation produced smooth activation profiles and consistent convergence without numerical artifacts.

\subsection{Adjustment of muscle parameters of an individual}

EMG-informed calibration resulted in parameter adjustments exceeding 10\% for three muscle groups: scapular deltoid, clavicular trapezius, and serratus anterior. The optimized maximum isometric force scalers were 1.16 for scapular deltoid, 1.02 for clavicular trapezius, and 1.17 for serratus anterior. The corresponding optimal fiber length scalers were 0.96, 0.74, and 0.92, respectively.

Across the validation performed on ADLs, calibration reduced RMSE between predicted muscle excitations and measured EMG for all three muscles. For lateral and posterior deltoid, RMSE changed from 0.126 to 0.122 (3.2\% reduction) and from 0.075 to 0.076 (1.3\% increase), respectively. For clavicular trapezius, RMSE decreased from 0.138 to 0.055 (60.1\% reduction). For serratus anterior, RMSE decreased from 0.255 to 0.189 (25.9\% reduction).

Muscle excitations obtained with original and adjusted parameters during the shelf-reaching task are shown in Fig.~\ref{fig,activation_validation}.

\subsection{Scapulohumeral rhythm prediction}

All simulations presented in this section were performed using EMG-informed adjusted muscle parameters. The RC-limited case simulation shown was simulated with $w_{GH}=10$. Results obtained with original (non-calibrated) muscle parameters are provided in the SM.

Fig. \ref{fig,kinematics_prediction} shows clavicular, scapular and humeral kinematics for healthy and RC-limited case across elevations in frontal, scapular and sagittal plane, respectively.

In the healthy condition, thoracohumeral elevation was accurately reproduced while scapular and clavicular kinematics emerged predictively from the optimization. Motion remained smooth across transitions between elevation planes. Phase-specific SHR values for all elevation planes are summarized in Table~\ref{tab,SHR values}.

In the RC-limited condition, scapulothoracic contribution was reduced. During frontal and sagittal plane elevations, peak scapular upward rotation decreased by approximately 10$^\circ$ and 12$^\circ$ relative to the healthy simulation, respectively. In scapular plane elevation, the reduction was more pronounced (approximately 18$^\circ$).

Fig.~\ref{fig,EMG_prediction} presents measured EMG signals together with predicted muscle excitations for healthy and RC-limited simulations. In the healthy condition, predicted activation profiles reproduced the general temporal patterns of the measured EMG. For sagittal plane elevation, infraspinatus excitation exceeded measured EMG amplitude, whereas lateral deltoid activation and clavicular trapezius were slightly lower than measured values in scapular and sagittal planes. Serratus anterior exceeded measured EMG in all elevation planes.

In both conditions, the GH reaction force vector remained centered within the glenoid throughout the motion, indicating maintained joint stability (Fig. \ref{fig,GH_stability}). However, the RC-limited simulation exhibited increased GH joint reaction force magnitude relative to the healthy condition. The increase was most pronounced during scapular elevation, where kinematic differences were largest.

\begin{table}[t]
\centering
\caption{Phase-specific scapulohumeral rhythm (SHR) for each elevation plane.}

\begin{tabular}{llcccc}
\hline
\textbf{Plane} & \textbf{Condition} 
& \textbf{start--30$^\circ$} 
& \textbf{30--60$^\circ$} 
& \textbf{60--90$^\circ$} 
& \textbf{start--90$^\circ$} \\
\hline
Frontal & Healthy      & 2.96 & 2.25 & 1.10  & 1.89  \\
        & RC-limited   & 5.44 & 2.77 & 1.85 & 2.89 \\
\hline
Scapular & Healthy     & 3.33 & 1.60 & 2.10 & 2.10  \\
         & RC-limited  & 4.88 & 4.61 & 2.28 & 5.17 \\
\hline
Sagittal & Healthy     & 2.88 & 2.05 & 2.39 & 2.29  \\
        & RC-limited   & 5.93 & 2.36 & 1.55 & 2.85 \\
\hline
\label{tab,SHR values}
\end{tabular}
\end{table}

\begin{figure}[t]
\centering
\centerline{\includegraphics[width=\columnwidth]{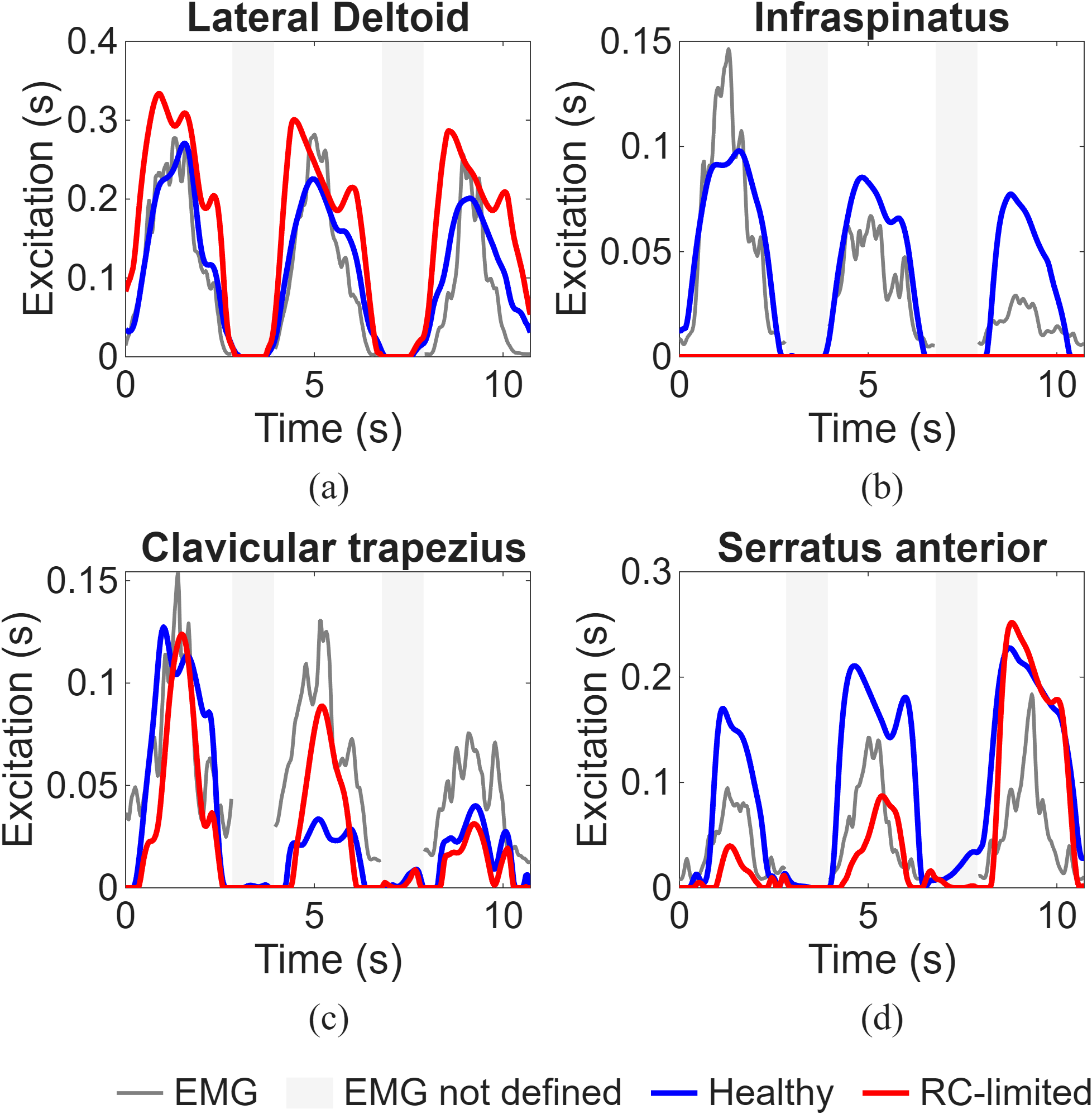}}
\caption{Muscle excitations during the healthy and RC-limited condition compared to the EMG signal (EMG is not defined during artificially added pauses).(a) Lateral deltoid exhibits higher excitations during RC-limited condition, whereas (c) clavicular trapezius and (d) serratus excitations were decreased. (b) Infraspinatus muscle was excluded in RC-limited simulation.}
\label{fig,EMG_prediction}
\end{figure}

\begin{figure}[t]
\centering
\centerline{\includegraphics[width=\columnwidth]{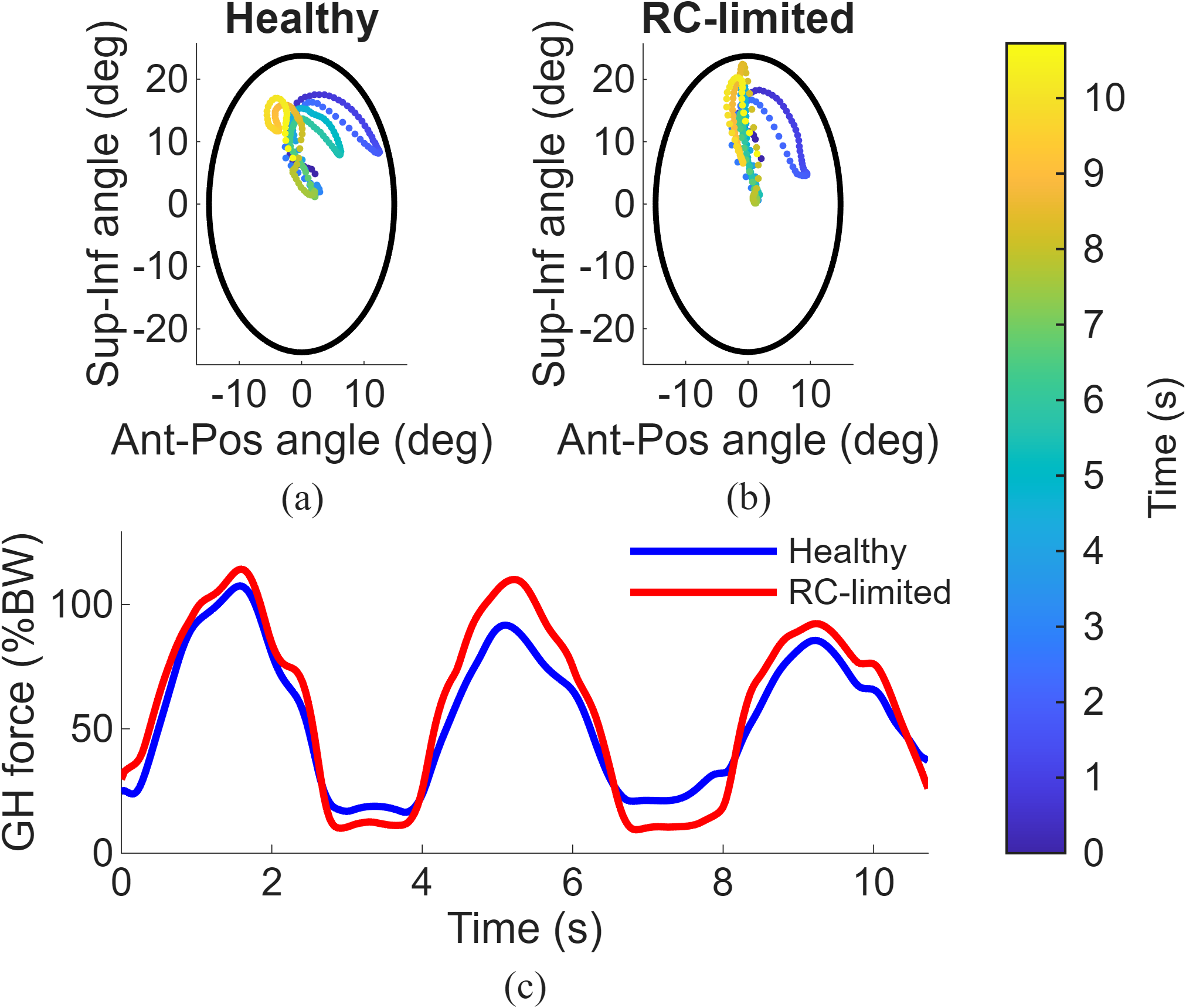}}
\caption{GH reaction force vector in the glenoid coordinate system for (a) healthy ($w_{GH}=2$) and (b) RC-limited simulation ($w_{GH}=10$). (c) GH reaction force is higher for the RC-limited simulation, particularly in the scapular and sagittal planes, where the kinematics discrepancies are largest.}
\label{fig,GH_stability}
\end{figure}

\subsection{Sensitivity to GH stability weighting}
\label{subsec,GH sensitivity}

A sensitivity analysis was performed to evaluate the influence of the GH stability weight on the RC-limited simulations. At the healthy stability weight, the RC-limited simulation already exhibited reduced scapular upward rotation by 8–15$^\circ$ across elevation planes relative to the healthy condition, indicating that the coordination shift is initiated by cuff removal itself. However, the GH reaction force vector temporarily exceeded the glenoid boundary during parts of the motion, indicating insufficient joint centering. Increasing the stability weight progressively restored joint centering while further altering the scapular kinematics. The value $w_{GH}=10$ was the minimum weight that maintained GH joint centering throughout the motion and was therefore used for the RC-limited simulations. Additional details are provided in the SM.

\section{Discussion}

The primary aim of this study was to develop and evaluate a computational quaternion-based framework for predictive simulation of scapulohumeral coordination. The following discussion therefore focuses on the methodological contributions — quaternion-based formulation, EMG-informed calibration, and predictive capability — and assesses whether the framework produces biomechanically plausible outputs. The RC–limited simulation serves as a demonstration of the framework's capability to investigate coordination redistribution under altered muscle conditions and should not be interpreted as a patient-specific RC pathology.

Phase-based analysis of scapulohumeral rhythm (SHR) demonstrated that the predictive framework reproduces physiologically plausible coordination patterns in healthy shoulder simulations. In the healthy case, total SHR values ranged approximately from 1.9–2.3 across all elevation planes, consistent with previously reported experimental measurements of SHR in healthy adults \cite{inman_observations_1996,mcclure_direct_2001,ludewig_alterations_2000}. Early elevation phases showed higher ratios due to minimal scapular rotation, which aligns with known phase-dependent variability of SHR \cite{mcquade_dynamic_1998}.

In the RC-limited simulation, SHR increased markedly during early and mid-elevation phases, indicating reduced scapular contribution and greater reliance on GH motion. This shift is mechanically consistent with the reduced stabilization capacity following RC limitation: at higher GH elevation angles, the deltoid force vectors are oriented more toward the glenoid center, allowing the muscle to contribute simultaneously to elevation and joint stabilization.

The predicted pattern of increased GH reliance in the absence of RC contribution shows qualitative similarities with trends observed by \cite{scibek_shoulder_2008}, who reported increased GH contribution in selected elevation phases following subacromial lidocaine injection in patients with full-thickness tears. However, a direct comparison is not appropriate: the present simulation represents complete exclusion of supraspinatus and infraspinatus to stress-test the framework, whereas participants in \cite{scibek_shoulder_2008} retained a partially functional cuff with structural tears. The similarity is therefore limited to the general direction of the coordination shift toward greater GH reliance when RC capacity is reduced rather than to the magnitude or phase-specific pattern of SHR changes. In contrast, chronic symptomatic RC pathology is typically associated with decreased SHR and increased scapulothoracic contribution \cite{ueda_comparison_2019,paletta_shoulder_1997}, reflecting long-term neuromuscular adaptations in the presence of pain. The divergence between the present predictions and chronic clinical findings suggests that pain-related motor adaptations are a critical determinant of shoulder coordination strategy and that their inclusion in the cost function may substantially alter predicted coordination patterns.

The RC–limited simulations demonstrated systematic changes in muscle activation consistent with the observed kinematic adaptations. Lateral deltoid activation increased across all planes, while activation of serratus anterior decreased in the frontal and scapular planes.

Sensitivity analysis showed that the predicted coordination pattern under cuff deficiency depended to some extent on the GH stability cost function weight. Increasing the stability weight progressively altered scapular kinematics during scapular and sagittal plane elevations, reflecting the trade-off between muscle activation minimization and joint stabilization. Higher stability weights increased the predicted activation of the remaining GH stabilizers, particularly teres minor and subscapularis, consistent with the greater demand for compressive joint loading in the absence of supraspinatus and infraspinatus.

EMG-informed calibration primarily affected serratus anterior and clavicular trapezius parameters, resulting in substantial reductions in RMS error between predicted excitations and measured EMG during validation tasks. The most pronounced improvement was observed for clavicular trapezius, where RMS error decreased by approximately 60\%. The substantial adjustment in optimal fiber length for this muscle suggests that its passive force–length properties strongly influence scapular posture and stabilization in the generic model. Shifting the operating length effectively increased passive force contribution reducing the need for elevated excitation to maintain clavicular position. Serratus anterior also showed meaningful improvement, although residual discrepancies remained. The magnitude of parameter adjustment was constrained by the regularization term to prevent overfitting and maintain physiologically plausible muscle properties. While further reductions in RMS error could potentially be achieved by decreasing regularization weights, this would risk overfitting of the model.

Compared to predictive simulations performed with generic parameters (SM), EMG-informed calibration reduced unrealistic clavicular depression at rest and improved overall neuromuscular consistency during SHR prediction. In addition to improving agreement with measured EMG, calibration brought clavicular kinematics closer to experimental observations during scapular plane elevation reported by \cite{mcclure_direct_2001}, who described approximately 2–6$^{\circ}$ of clavicular elevation and -17 to -25$^{\circ}$ of retraction in healthy subjects. Following calibration, the present model predicted approximately 5–12$^{\circ}$ of elevation and -20 to -32$^{\circ}$ of retraction, whereas simulations with generic parameters produced broader and less physiologically consistent ranges (-8 to 10$^{\circ}$ elevation and up to -45$^{\circ}$ retraction). This indicates that EMG-informed parameter adjustment enables fully predictive scapuloclavicular coordination and reduces the need for additional kinematic constraints to manage the high dimensionality of the system. In contrast to a recent framework that prescribes sternoclavicular elevation and enforces scapulothoracic contacts to restrict the solution space \cite{russell_shoulder_2026}, the present approach allows coordination to emerge from musculoskeletal mechanics and calibrated muscle properties.

Despite these improvements, predicted infraspinatus excitation exceeded measured EMG during sagittal plane elevation, while clavicular trapezius excitation was lower than measured EMG in scapular and sagittal planes. These discrepancies may reflect the simplified GH stability representation and the increased passive force contribution following optimal fiber length adjustment, respectively.

In the healthy simulation, peak magnitude of GH reaction force reached 107.5\% BW during frontal plane elevation and 85.5\% BW during sagittal elevation. Instrumented implant data from the Orthoload database report mean peak GH contact forces of 85.3 ± 26.4\% BW (frontal plane, n = 6) and 71.0 ± 9.0\% BW (sagittal plane, n = 3). Predicted GH reaction forces in the healthy simulation were moderately higher than the mean values reported in the instrumented implant data, but remained within one standard deviation in the frontal plane data. Differences in predicted GH loading may arise from variations in version and inclination angles across participants as well as the absence of passive ligament contributions and simplified joint geometry. Despite maintained joint centering in both conditions, the RC-limited simulation exhibited increased GH joint reaction compared to the healthy simulation, which may reflect altered force redistribution required to preserve stability under reduced cuff contribution.

From a methodological perspective, the unit-quaternion representation was fundamental to the framework's numerical robustness, successfully eliminating the kinematic singularities and associated activation artifacts observed in the Euler-angle formulation. To ensure unit length of quaternions throughout the simulation, a hybrid constraint strategy was employed: the unit-norm condition was strictly enforced via a hard algebraic equality constraint at the initial node ($t_0$), while subsequent nodes were regulated using a soft-penalty term in the objective function. Post-hoc verification confirmed that this strategy achieved an RMSE of unit length of quaternion on the order of $10^{-14}$ or lower across all collocation nodes.

Several limitations should be acknowledged. The GH joint was modeled as an ideal spherical joint without translational degrees of freedom or explicit articular contact constraints. In reality, GH motion involves coupled rotation and translation influenced by bone geometry, capsuloligamentous structures, and soft tissue contact mechanics. These anatomical constraints may limit extreme rotational strategies that remain mechanically feasible in the simplified representation. Passive GH ligaments were not explicitly modeled, and joint stability was approximated using a simplified elliptical glenoid representation.

While the present results demonstrate that the framework produces coordination patterns consistent with experimentally observed biomechanical trends, the simulations were performed on a single participant as a methodological proof-of-concept. Translation toward clinical application would require evaluation across a larger cohort to assess inter-subject variability, incorporation of subject-specific glenoid morphology, and investigation of levels of RC pathologies including partial tears or isolated supraspinatus deficiency.

Future developments may also incorporate detailed joint contact modeling, passive ligament mechanics, and pain-related objective terms to further enhance predictive accuracy.

Furthermore, thoracohumeral elevation was prescribed to isolate scapular and clavicular coordination responses under altered muscle conditions. While this enabled controlled analysis of SHR redistribution, it prevented full upper-limb motion from emerging solely from task-level objectives. Future work may omit this constraint by prescribing only initial conditions and endpoint targets, allowing complete movement trajectories to be determined by optimal control. Such extensions would enable investigation of coordination strategies during dynamic tasks, including overhead throwing, wheelchair propulsion, or other high-demand activities.

\section{Conclusion}

This study presented an EMG-informed predictive optimal control framework for investigating scapulohumeral coordination. The quaternion-based formulation improved numerical robustness and eliminated singularity-related artifacts observed in Euler-angles simulations. EMG-informed adjustment of muscle–tendon parameters enhanced agreement with measured activation patterns and improved emergent claviculothoracic and scapulothoracic coordination. The predictive simulations reproduced physiologically plausible SHR in healthy conditions and demonstrated mechanically consistent redistribution of motion and muscle activation following simulated RC deficiency.

Together, these findings highlight the potential of predictive optimal control approaches to investigate coordination strategies and joint loading adaptations in the shoulder complex, providing a foundation for future subject-specific and task-driven simulations.

\section{Acknowledgment}
This research was supported by the Czech Ministry of Education, Youth and Sports project No. CZ.02.01.01/00/23\_020/0008512 and Czech Grant Agency project No. 23-06920S.

\clearpage
\section*{REFERENCES}
\bibliographystyle{IEEEtran}
\bibliography{ShoulderBiomechanics}
\end{document}


\title{Quaternion-Based Predictive Framework for Scapulohumeral Coordination}
\author{Ondrej Zoufaly, Edward K. Chadwick, Dimitra Blana, Matej Daniel}

\maketitle

\renewcommand{\thesection}{S\Roman{section}} 
\renewcommand{\thefigure}{S\arabic{figure}}  
\renewcommand{\thetable}{S\Roman{table}} 
\setcounter{section}{0}
\setcounter{figure}{0}
\setcounter{table}{0}

\section{Muscle--Tendon Model}

Muscle forces were computed using a Hill-type muscle--tendon model consisting of an activation-dependent contractile element and nonlinear passive elastic properties. The formulation follows the structure of the classical Hill-type muscle model described by \cite{thelen_adjustment_2003} and incorporates the smooth force--velocity formulation proposed by \cite{de_groote_evaluation_2016} to ensure full differentiability within the optimal control framework.

Muscle fiber length $l_m$ was computed from the musculotendon length $l_{mt}$ and tendon slack length $l_{slack}$

\begin{equation}
l_m = l_{mt} - l_{slack}.
\end{equation}

This simplified representation assumes a rigid tendon without explicit tendon compliance, allowing the contractile element length to be directly derived from the musculotendon length.

\subsection{Active Force--Length Relationship}

The normalized active force--length relationship was represented using a Gaussian function centered around the optimal fiber length $l_{ce}^{opt}$

\begin{equation}
f_{lce} =
\exp
\left(
-\frac{(l_m/l_{ce}^{opt}-1)^2}{f_{gauss}}
\right).
\end{equation}

\subsection{Passive Force--Length Relationship}

Passive muscle force was modeled using an exponential relationship

\begin{equation}
f_{pe} =
\frac{\exp
\left(
k_{pe}\frac{(l_m/l_{ce}^{opt}-1)}{\varepsilon_{m0}}
\right)-1}
{\exp(k_{pe})-1}.
\end{equation}

\subsection{Force--Velocity Relationship}

Muscle force dependence on contraction velocity was modeled using the continuous logarithmic formulation described by De Groote et al.~\cite{de_groote_evaluation_2016}. The normalized contraction velocity was defined as

\begin{equation}
v_{norm} = \frac{v_{ce}}{v_{max}},
\end{equation}

where $v_{ce}$ is the contractile element velocity and $v_{max}$ is the maximum shortening velocity normalized by optimal fiber length.

The normalized force--velocity relationship was computed as

\begin{equation}
f_{vce} =
d_1 \ln
\left(
d_2 v_{norm} + d_3 +
\sqrt{(d_2 v_{norm} + d_3)^2 + 1}
\right)
+ d_4.
\end{equation}

\subsection{Muscle Force}

Total muscle force was computed as

\begin{equation}
F =
\left(
f_{lce} f_{vce} a + f_{pe}
\right) F_{max},
\end{equation}

where $a$ is muscle activation and $F_{max}$ is the maximum isometric force.

All parameter values used in the muscle model are summarized in Table~\ref{tab:muscle_parameters}.

\begin{table}[h]
\centering
\caption{Parameters used in the muscle--tendon model.}
\label{tab:muscle_parameters}
\begin{tabular}{lll}
\hline
Parameter & Description & Value \\
\hline
$f_{gauss}$ & Width of active force--length curve & 0.25 \\
$k_{pe}$ & Passive force exponential stiffness & 5 \\
$\varepsilon_{m0}$ & Passive strain at unity force & 0.6 \\
$d_1$ & Force--velocity coefficient & -0.318 \\
$d_2$ & Force--velocity coefficient & -8.149 \\
$d_3$ & Force--velocity coefficient & -0.374 \\
$d_4$ & Force--velocity offset & 0.886 \\
$v_{max}$ & Maximum shortening velocity (normalized) & $10\times l_{ce}^{opt}$ \\
\hline
\end{tabular}
\end{table}

\subsection{Muscle activation dynamics}

Muscle activation dynamics was modeled using a first-order differential equation \cite{chadwick_real-time_2014}:

\begin{equation}\label{eq,act_dynamics}
\frac{da}{dt} =
\left(
\frac{e}{T_{act}} + \frac{1-e}{T_{deact}}
\right)(e-a),
\end{equation}

where $a$ denotes muscle activation, $e$ denotes neural excitation input, and $T_{act}$ and $T_{deact}$ represent activation and deactivation time constants, respectively.

\subsection{Scaling of the trajectory tracking weight for quaternion formulation}

Trajectory tracking was implemented using a quadratic penalty on the error between simulated and experimentally measured joint orientations. Because the Euler-angle and quaternion formulations use different orientation representations, the magnitude of the tracking residual differs between the two formulations and requires appropriate scaling of the tracking weight.

In the Euler-angle formulation, orientation is represented by three generalized coordinates
\[
\boldsymbol{\theta} = [\theta_1,\theta_2,\theta_3]^T,
\]
and the tracking cost is computed as

\begin{equation}
J_{track}^{Eul} = w_{track}^{Eul} \|\boldsymbol{\theta}-\boldsymbol{\theta}_{ref}\|^2 .
\end{equation}

In the quaternion formulation, orientation error was computed using the relative quaternion between the simulated orientation \(q\) and the reference orientation \(q_{ref}\). The relative quaternion was obtained through quaternion multiplication

\begin{equation}
\textbf{Q}_{err} = \textbf{Q}\otimes \textbf{Q}^{-1}_{exp}
\end{equation}

For a unit quaternion
\[
\textbf{Q}_{err} = [Q_0,\mathbf{Q_v}],
\]
the vector part \(\mathbf{Q_v}\) represents the axis–angle rotation error. The tracking cost was therefore defined using the squared magnitude of the imaginary component

\begin{equation}
J_{traj}^{Q} = w_{traj}^{Q} \|\mathbf{Q_v}\|^2 .
\end{equation}

For small rotational errors, the relative quaternion can be approximated as

\[
\textbf{Q}_{err} \approx
\left[
1,
\frac{\delta\boldsymbol{\phi}}{2}
\right],
\]

where \(\delta\boldsymbol{\phi}=[\delta\phi_x,\delta\phi_y,\delta\phi_z]^T\) denotes the small-angle rotation error vector. The squared magnitude of the imaginary component therefore becomes

\begin{equation}
\|\mathbf{Q_v}\|^2 \approx
\frac{1}{4}
\|\delta\boldsymbol{\phi}\|^2 .
\end{equation}

In contrast, the Euler-angle formulation penalizes the squared angular deviation directly

\begin{equation}
\|\boldsymbol{\theta}-\boldsymbol{\theta}_{ref}\|^2
\approx
\|\delta\boldsymbol{\phi}\|^2 .
\end{equation}

Thus, for small orientation errors the quaternion tracking residual is approximately four times smaller than the corresponding Euler-angle residual. To maintain comparable influence of the tracking term within the objective function, the quaternion tracking weight was therefore scaled approximately fourfold relative to the Euler formulation:

\[
w_Q \approx 4 w_E .
\]

It should be noted that this scaling is only approximate. Euler angles describe three independent rotational coordinates, whereas quaternions encode orientation as a single axis–angle rotation within a four-parameter unit vector representation. The equivalence therefore holds only locally for small rotations, and the factor of four serves as a practical normalization to obtain comparable tracking strength between formulations.

\begin{table*}[t]
\centering
\caption{EMG-informed muscle parameter adjustment. 
Each virtual muscle group was assigned a single parameter scaler, 
adjusted to improve consistency between simulated muscle activations 
and experimentally measured EMG signals.}
\label{tab:emg_groups}
\begin{tabular}{lll}
\hline
\textbf{Virtual muscle group} & \textbf{Muscle elements used to match the EMG} & \textbf{EMG reference} \\
\hline
Clavicular trapezius 
& \texttt{trap\_clav\_1} 
& Clavicular trapezius \\

Clavicular deltoid 
& \texttt{delt\_clav\_1}, \texttt{delt\_clav\_2} 
& Anterior deltoid \\

Scapular deltoid
& \texttt{delt\_scap\_8}, \texttt{delt\_scap\_9}, \texttt{delt\_scap\_10}, \texttt{delt\_scap\_11} 
& Lateral deltoid \\

& \texttt{delt\_scap\_3}, \texttt{delt\_scap\_4}, \texttt{delt\_scap\_5}
& Posterior deltoid \\

Infraspinatus 
& \texttt{infra\_2}, \texttt{infra\_3}, \texttt{infra\_4} 
& Infraspinatus \\

Scapular trapezius
& \texttt{trapscap\_1}, \texttt{trapscap\_2} 
& Upper scapular trapezius \\

& \texttt{trapscap\_5}, \texttt{trapscap\_6} 
& Scapular trapezius at level of scapular spine \\

Serratus anterior 
& \texttt{serr\_ant\_2}, \texttt{serr\_ant\_3}, \texttt{serr\_ant\_4} 
& Serratus anterior at level of angulus inferior \\
\hline
\end{tabular}
\end{table*}

\begin{figure*}[t]
\centering
\centerline{\includegraphics[width=\textwidth]{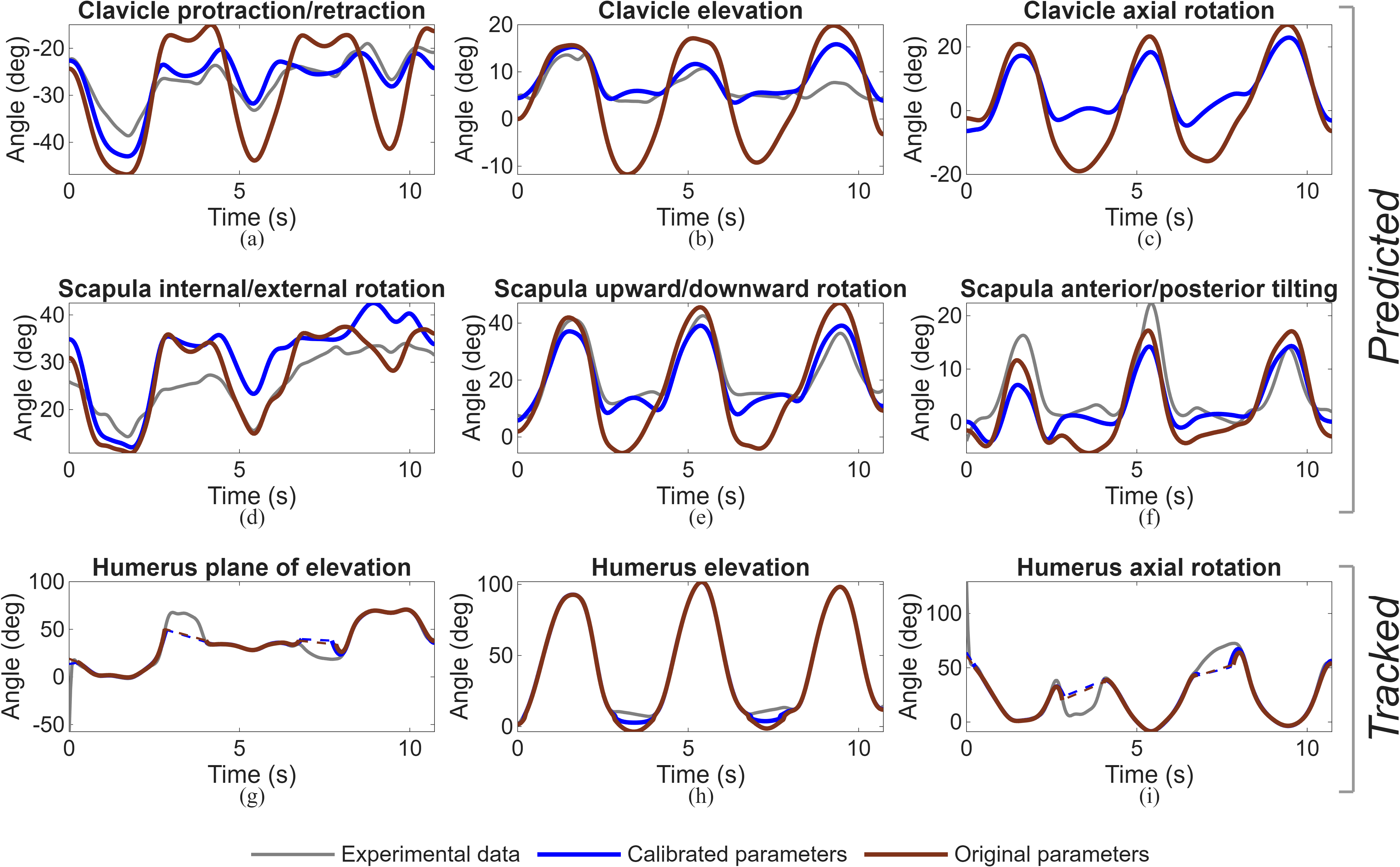}}
\caption{SHR prediction using original and adjusted muscle parameters. The prediction using original parameters shows high clavicular depression and scapular downward rotation during arm neutral position.}
\label{fig,kinematics_prediction_origvsadjust}
\end{figure*}

\section{Muscle elements subjected to calibration}

Muscle elements (with corresponding virtual muscle group and EMG reference) used for muscle parameters calibration and EMG comparison are shown in Table \ref{tab:emg_groups}. If multiple elements were selected, the average of them was calculated.

\section{SHR prediction using original vs adjusted parameters}

SHR prediction using original parameters show excessive clavicular depression and scapular downward rotation due to insufficient force capacity of clavicular trapezius muscles in neutral position (Fig. \ref{fig,kinematics_prediction_origvsadjust}) At 90$^\circ$ of thoracohumeral elevation, scapular upward rotation closely matched the values predicted by the prediction using adjusted paremeters. This alignment ensures the deltoid muscles operate near their optimal fiber lengths.

\section{GH weight sensitivity analysis}

Sensitivity analysis was performed was the RC-limited case starting with $w_{GH}$ of 2 (=healthy case) to 10. Fig. \ref{fig,GHstab_sanalysis} shows GH joint reaction vector pointing more toward the glenoid center gradually as the weight increases until the GH get stable for $w_{GH}=10$. During the weight increase there is also a gradual alteration of the kinematics (Fig.~\ref{fig,kinematics_sanalysis}), primarily affecting the scapular upward/downward rotation.

\begin{figure*}[t]
\centering
\centerline{\includegraphics[width=\textwidth]{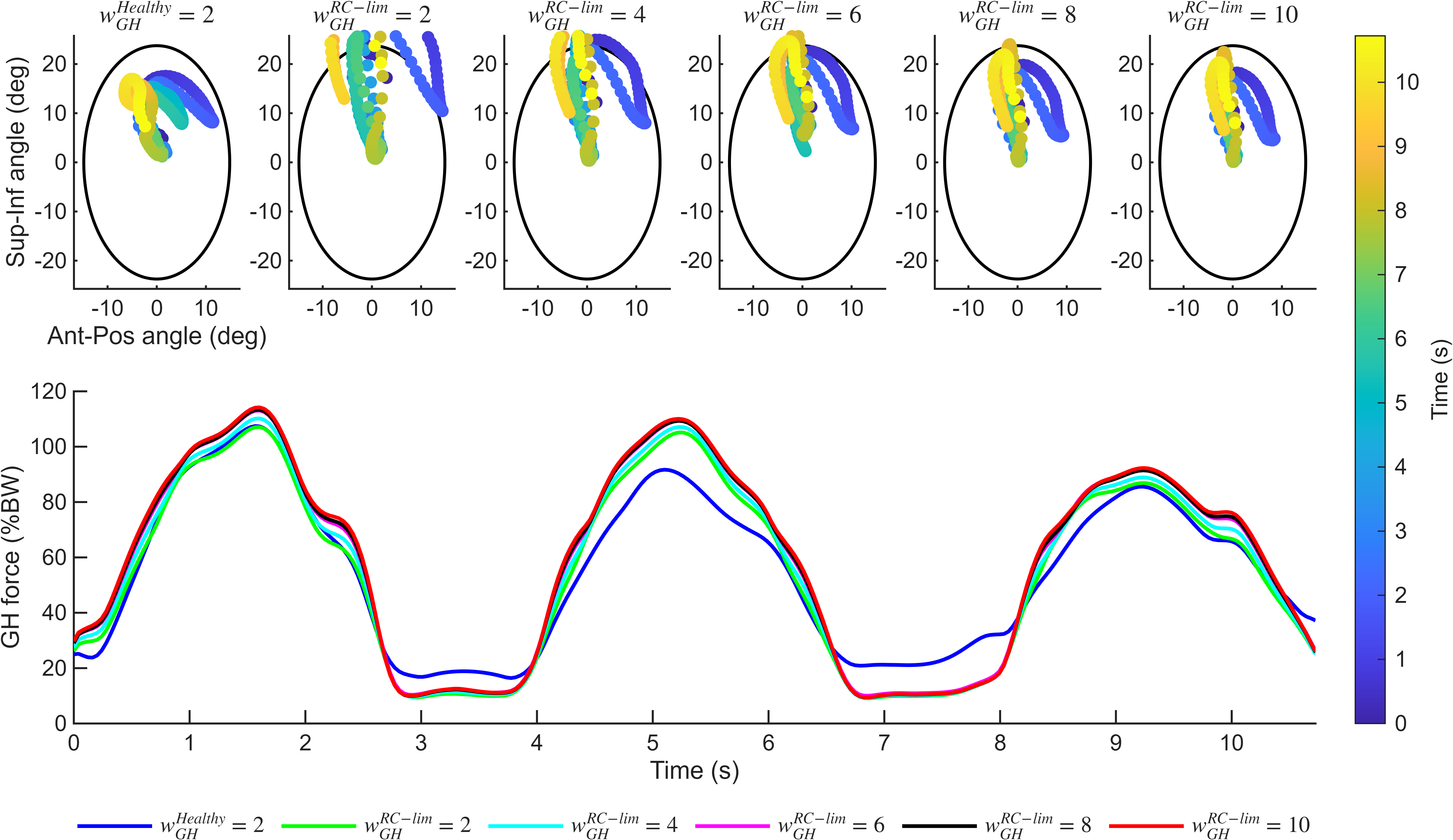}}
\caption{Sensitivity analysis for RC-limited case performed on the GH stability weight. For the value used for the healthy case ($w^{RC-lim}_{GH} = w^{Healthy}_{GH} = 2$), GH joint is not stable. $w^{RC-lim}_{GH}$ was increased until the joint centering became sufficient.}
\label{fig,GHstab_sanalysis}
\end{figure*}

\begin{figure*}[t]
\centering
\centerline{\includegraphics[width=\textwidth]{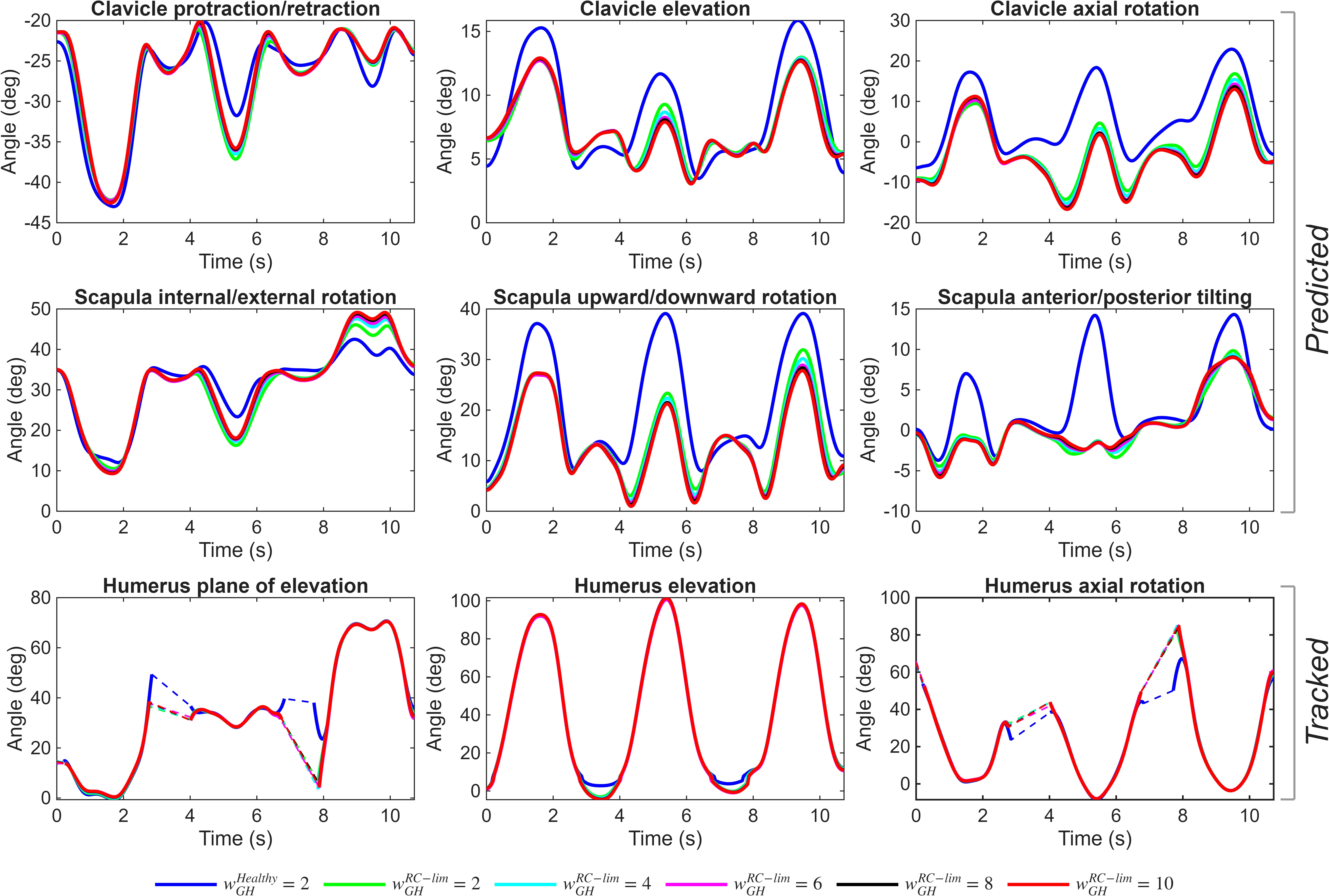}}
\caption{Sensitivity analysis of the RC-limited case for different GH stability weights. The kinematics alteration in RC-limited case is already substantial for $w^{RC-lim}_{GH} = w^{Healthy}_{GH} = 2$. Further joint centering alters scapular kinematics slightly.}
\label{fig,kinematics_sanalysis}
\end{figure*}

\section*{REFERENCES}
\bibliographystyle{IEEEtran}
\bibliography{ShoulderBiomechanics}